\documentclass[12pt]{article}
\usepackage{graphicx}
\usepackage{xspace}
\usepackage{amsmath}
\usepackage{hyperref}
\newcommand\pubnumber{SNSN-323-63}
\newcommand\pubdate{November 15, 2014}

\def\cern{European Organisation for Nuclear Research\\
	CH1211, Gen\`{e}ve 23, Switzerland}
\def\support{\footnote{On behalf of LHCb and CMS collaborations.}}

\def\lhcb{LHCb\xspace}
\def\cms{CMS\xspace}
\def\bds{\ensuremath{B_{(s)}^0}\xspace}
\def\bd{\ensuremath{B^0}\xspace}
\def\bs{\ensuremath{B_{s}^0}\xspace}

\def\bdsmumu{\ensuremath{B_{(s)}^0 \rightarrow \mu^+\mu^-}\xspace}
\def\bdmumu{\ensuremath{B^0 \rightarrow \mu^+\mu^-}\xspace}
\def\bsmumu{\ensuremath{B_{s}^0 \rightarrow \mu^+\mu^-}\xspace}

\def\fb{\ensuremath{\,\mathrm{fb}^{-1}}\xspace}

\def\Bhhprime{\ensuremath{B^0_{(s)}\to h^+h^{'-}}\xspace}

\def\bdkpi{\ensuremath{B^0\to K^+ \pi^-}\xspace}
\def\bujpsik{\ensuremath{B^+\to J/\psi K^+}\xspace}

\def\Bhmumu{\ensuremath{B \to h \mu\mu}\xspace}
\def\Bhmunu{\ensuremath{B \to h \mu\nu}\xspace}
\def\Lbpmunu{\ensuremath{\Lambda^0_b \to p \mu^-\bar{\nu}}\xspace}

\def\fb{\ensuremath{\,\mathrm{fb}^{-1}}\xspace}

\def\mmumu{\ensuremath{m_{\mu\mu}}\xspace}

\def\fsfd{\ensuremath{f_s/f_d}\xspace}

\def\br[#1]#2{\ensuremath{\mathcal{B}(#1) = #2}\xspace}
\def\R[#1]#2{\ensuremath{\mathcal{R}^{#1} = #2}\xspace}
\newcommand{\BRof}[1]{\ensuremath{{\cal B}(#1)}\xspace}
\newcommand{\BRSMof}[1]{\ensuremath{{\cal B^{\mathrm{SM}}}(#1)}\xspace}
\def\RB{\ensuremath{{\mathcal{R}}}\xspace}

\newcommand{\RBSM}{\ensuremath{0.0295^{+0.0028}_{-0.0025}}\xspace} 
\def\bsmumumeas{\ensuremath{\left( 2.8 \,^{+0.7}_{-0.6} \right) \times 10^{-9}}\xspace}
\def\bdmumumeas{\ensuremath{\left( 3.9 \,^{+1.6}_{-1.4} \right) \times 10^{-10}}\xspace}

\def\Title#1{\begin{center} {\Large #1 } \end{center}}
\def\Author#1{\begin{center}{ \sc #1} \end{center}}
\def\Address#1{\begin{center}{ \it #1} \end{center}}

\newcommand\pubblock{\rightline{\begin{tabular}{l} \pubnumber\\
         \pubdate  \end{tabular}}}
\newenvironment{Abstract}{\begin{quotation}  }{\end{quotation}}
\newenvironment{Presented}{\begin{quotation} \begin{center} 
             PRESENTED AT\end{center}\bigskip 
      \begin{center}\begin{large}}{\end{large}\end{center} \end{quotation}}

\def\beq{\begin{equation}}
\def\eeq#1{\label{#1}\end{equation}}
\def\eeqn{\end{equation}}

\def\beqa{\begin{eqnarray}}
\def\eeqa#1{\label{#1}\end{eqnarray}}
\def\eeqan{\end{eqnarray}}

\let\bar=\overbar

\def\Dslash{\not{\hbox{\kern-4pt $D$}}}
\def\dslash{\not{\hbox{\kern-2pt $\del$}}}

\def\msb{{\bar{\ssstyle M \kern -1pt S}}}

\begin{document}
\begin{titlepage}
\pubblock

\vfill
\Title{$B^0_{s}\rightarrow \mu^+\mu^-$ at LHC}
\vfill
\Author{ Flavio Archilli\support}
\Address{\cern}
\vfill
\begin{Abstract}
Rare leptonic decays of $B_{(s)}^0$ mesons are sensitive probes of New Physics effects. A combination combination of the \cms and \lhcb analyses on the search for the rare decays \bsmumu and \bdmumu is presented. The branching fractions of \bsmumu and \bdmumu are measured to be \br[\bsmumu]{\bsmumumeas} and \br[\bdmumu]{\bdmumumeas}, respectively. A statistical significances of $6.2\,\sigma$ is evaluated for \bsmumu from the Wilks' theorem while a significance of $3.0\, \sigma$ is measured for \bdmumu from the Feldman-Cousins procedure.
  
\end{Abstract}
\vfill
\begin{Presented}
Presented at the 8th International Workshop on the CKM Unitarity Triangle (CKM 2014), Vienna, Austria, September 8-12, 2014
\end{Presented}
\vfill
\end{titlepage}
\def\thefootnote{\fnsymbol{footnote}}
\setcounter{footnote}{0}

\section{Introduction}
Limits on the rare \bdsmumu decays Branching Fractions (BF) are one of the most promising ways to constrain New Physics (NP) models. 
These decays are highly suppressed in the Standard Model (SM), because they are flavour changing neutral current processes, that occur through $Z$ penguin diagrams or $W$-box diagrams. Moreover, the helicity suppression of axial vector terms makes these decays par\-tic\-u\-lar\-ly sensitive to NP scalar and pseudoscalar contributions, such as extra Higgs doublets, that can raise their BF with respect to SM expectations.
The untagged time-integrated SM predictions for these decays are~\cite{bsmumu:bobeth}:
\begin{eqnarray}
\mathcal{B}(\bsmumu)_{SM} = (3.66\pm0.23)\times 10^{-9} \nonumber \,,\\
\mathcal{B}(\bdmumu)_{SM} = (1.06\pm0.09)\times 10^{-10} \nonumber \,,
\end{eqnarray}
which use the latest combined value for the top mass from LHC and Tevatron experiments~\cite{bsmumu:topmass}.
Moreover, the ratio \RB of the BFs of these two modes proves to be powerful to discriminate among models beyond the SM (BSM). This ratio is precisely predicted in the SM to be:
\begin{eqnarray}
\RB = \frac{\BRof{\bdmumu}}{\BRof{\bsmumu}}   
= \frac{\tau_{\bd}}{1/\Gamma_H^{s}} \left(\frac{f_{\bd}}{f_{\bs}}\right)^2 
 \left|\frac{V_{td}}{V_{ts}}\right|^2 \tfrac{M_{\bd} \sqrt{1 - \tfrac{4 
 m_{\mu}^2}{M_{\bd}^2}}}{M_{\bs} \sqrt{1 - \tfrac{4 m_{\mu}^2}{M_{\bs}^2}}}
= \RBSM\quad 
\end{eqnarray}
where $\tau_{\bd}$ and $1/\Gamma_H^{s}$ are the lifetimes of the \bd and of the heavy mass eigenstate 
of the \bs; $M_{\bds}$ is the mass and $f_{\bds}$ is the decay constant of the \bds meson; $V_{td}$ and $V_{ts}$ are the elements of the CKM matrix and $m_{\mu}$ is the mass of the muon. In BSM models with minimal flavour violation property this quantity is predicted to be equal to the SM ratio.

The LHCb collaboration has reported the first evidence of the \bsmumu decay with a $3.5\,\sigma$ 
significance~\cite{bsmumu:LHCbevidence} in 2012 using 2\fb\ collected during the first two years of data taking. In 2013, CMS and LHCb presented their updated results based on 25\fb and 3\fb, respectively~\cite{bsmumu:LHCb4sigma}~\cite{bsmumu:CMS4sigma}. 
The two measurements are in good agreement with each other, and have comparable precisions; however, none of them is precise enough to claim the first observation of the \bsmumu decay.

A na\"ive combination of \lhcb and \cms results was presented during the European Physical Society Conference on High Energy Physics in 2013~\cite{bsmumu:naivecomb}. The result was:
\begin{eqnarray}
\mathcal{B}(\bsmumu)_{SM} = (2.9\pm0.7)\times 10^{-9} \nonumber \,,\\
\mathcal{B}(\bdmumu)_{SM} = (3.6^{+1.6}_{-1.4})\times 10^{-10} \nonumber \,.
\end{eqnarray}
Despite they represent the most precise measurements on the rare decays \bdsmumu, 
no accurate attempt was made to take into account for all the correlations arising from the common physical quantities, and the statistical significance was not provided.
In these proceedings a combination of the results based on a simultaneous fit to the two datasets is presented. This fit correctly takes into account correlations between the input parameters.

\section{Analyses}
The \cms and \lhcb experiments have very similar analysis strategies. \bdsmumu candidates are selected as two oppositely charged tracks. A soft preselection is applied   
in order to reduce the background while keeping high the efficiency on the signal.
After this selection, the surviving backgrounds are mainly due to random combinations of muons generated in semileptonic $B$ decays (combinatorial background), semileptonic decays, such as \Bhmunu, \Bhmumu and \Lbpmunu, and \Bhhprime decays (peaking background) where hadrons are misidentified as muons. 
Further separation between signal and background is achieved exploiting the power of a multivariate classifier trained using the TMVA~\cite{bsmumu:tmva} framework. The classification of the events is done using the dimuon invariant mass \mmumu and the multivariate classifier output. \cms further classifies the candidates as ``barrel'', with both muons having a pseudorapidity $\vert \eta \vert < 1.4$ and ``endcap'', with at least one muon having $\vert \eta \vert > 1.4$. 
The multivariate classifier is, for both experiments, a boosted decision tree, BDT, and it is trained using kinematic and 
geometrical variables. 
The calibration of the dimuon mass \mmumu is performed using the dimuon resonances and, for \lhcb, also by using the \Bhhprime decays.
For both analyses the \bdsmumu yield is normalised with respect to the \bujpsik yield, taking into account the hadronisation fractions of a $b$ quark to \bs and \bd mesons measured by the \lhcb experiment~\cite{bsmumu:prefsfd}~\cite{bsmumu:fsfd}~\cite{bsmumu:fsfdUpdate}. \lhcb collaboration also used \bdkpi decay as a normalisation channel.
%%%%%%%%%%%%%%%

Some changes were made to harmonise the analyses: the \Lbpmunu background source was included in the nominal fit, with an updated BF and an updated Monte Carlo simulation in order to include a more realistic model for the properties of the decay and the lifetime bias~\cite{bsmumu:lifetime} correction on the signal PDF was applied to the \cms analysis, too. 

A simultaneous fit is performed to evaluate the BFs of the \bsmumu and \bdmumu decays. The two datasets from \cms and \lhcb analyses are used together in a single combined experiment. 
A simultaneous unbinned extended maximum likelihood fit is performed to the invariant mass spectrum in 20 categories for the two experiments, 8 categories for \lhcb and 12 categories for \cms.
In each category the mass spectrum is described as the sum of the PDF of each background source and the two signals. The parameters shared between the PDFs in the two experiments are: the BFs of the two signals decays \BRof{\bsmumu} and \BRof{\bdmumu}, the BF of the common normalisation channel \BRof{\bujpsik} and the ratio of the hadronisation fractions \fsfd.
Assuming the SM BFs, $94\pm7$ \bsmumu events and $10.5\pm0.6$ \bdmumu events are expected in the full dataset.

\section{Results}

\begin{figure}[!h]
\centering
\includegraphics[height=2in]{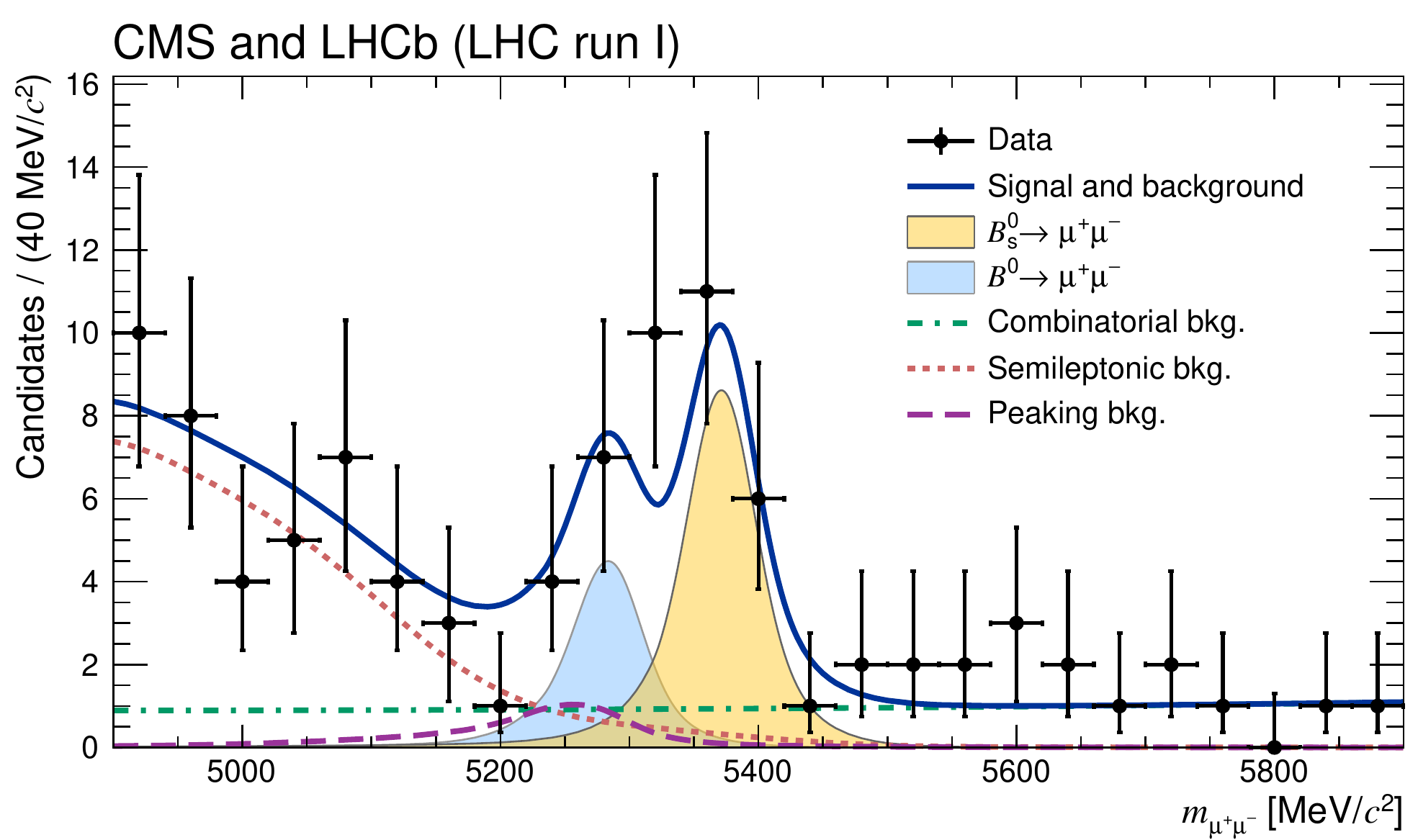}
\caption{Dimuon mass distribution projection for the best six categories ranked accordingly to S/(S+B) (see text) is superimposed, as well as the result of the simultaneous fit (solid blue) and each fit component: \bsmumu (yellow shaded); \bdmumu (light-blue shaded); combinatorial background (dash-dotted green); semileptonic backgrounds (dotted pink); peaking background (dashed violet).}
\label{fig1}
\end{figure}

\begin{figure}[!h]
\centering
\includegraphics[height=2in]{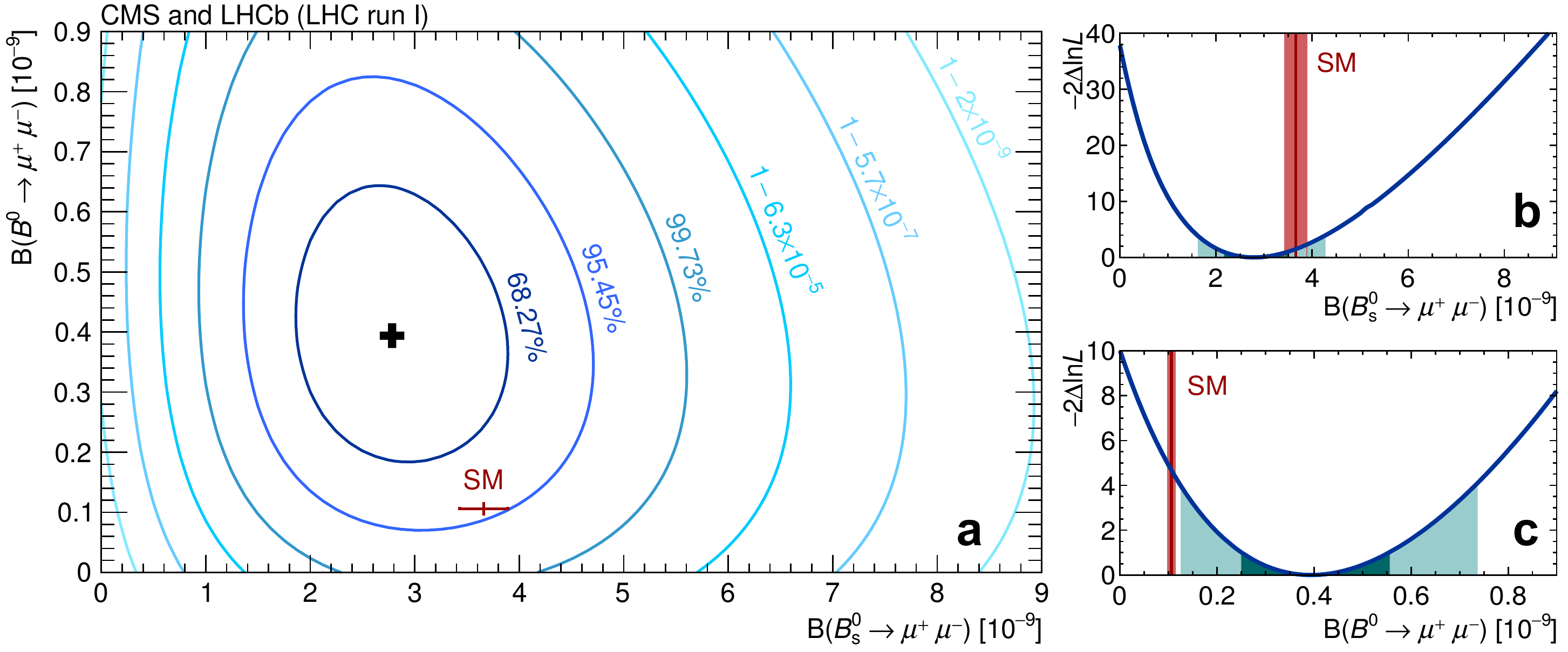}
\caption{Likelihood contours for \BRof\bsmumu as a function of \BRof\bdmumu (left). The SM expectation is reported as a red cross. Likelihood test-statistics ($-2\Delta \mathrm{ln}L$) profile for \BRof\bsmumu (top-right) and \BRof\bdmumu (bottom-right). The dark and light (cadet blue) areas define the $\pm 1\,\sigma$ and $\pm 2\,\sigma$ confidence intervals, respectively. The SM expectation and its uncertainty for each BF is denoted with a vertical (red) band.}
\label{fig2}
\end{figure}

The results of the simultaneous fit for the signal BFs are~\cite{bsmumu:nature}:
\begin{eqnarray}
\br[\bsmumu]{\bsmumumeas} \,, \nonumber \\
\br[\bdmumu]{\bdmumumeas} \,, \nonumber
\end{eqnarray}
where the uncertainties include both statistical and systematic errors.
In Fig.~\ref{fig1} the dimuon mass distribution is shown for the events falling in the best six categories, selected through S/(S+B) values, where S and B are the signal and the background yields, respectively, expected under the \bs mass peak assuming the SM BFs. The statistical significances, evaluated using the Wilks' theorem, are $6.2\,\sigma$ and $3.2\,\sigma$ for \bsmumu and \bdmumu respectively. The expected significances assuming the SM BFs are $7.4\,\sigma$ and $0.8\,\sigma$ for \bs and \bd channels respectively.
%%%%%%%%%%%%%%%%%%%%%%%%%%%%%%%%%%%%%%%%%%%%%%%%%%%%%%%%%%%%%%%%%%%%%%%%%
%%%%%%%%%%%%%%%%%%%%%%%%%%%%%%%%%%%%%%%%%%%%%%%%%%%%%%%%%%%%%%%%%%%%%%%%%%%
Since the Wilks' theorem shows a \bdmumu signal significance slightly above the $3\,\sigma$ level and moreover assumes asymptotic behavior close to the null hypothesis, a Feldman-Cousins~\cite{bsmumu:fc} based method has been also used for the \bdmumu mode. A statistical significance of $3.0\,\sigma$ is measured in this case. The Feldman-Cousins method confidence intervals at $\pm 1\,\sigma$ and $\pm 2\,\sigma$ are evaluated to be $[2.5,5.6]\times10^{-10}$ and $[1.4,7.4]\times10^{-10}$, respectively.
In Fig.~\ref{fig2} the likelihood contours for \BRof\bsmumu as a function of \BRof\bdmumu are shown. In the same figure the likelihood profile for each signal mode is displayed.
%%%%%%%%%%%%%%%%%%%%%%%%%%%%%%%%%%%%%%%%%%%%%%%%%%%%%%%%%%%%%%%%%%%%%%%%%
%%%%%%%%%%%%%%%%%%%%%%%%%%%%%%%%%%%%%%%%%%%%%%%%%%%%%%%%%%%%%%%%%%%%%%%%%%%

A simultaneous fit to the ratios of the BFs relative to their SM expectation values is also performed to evaluate the compatibility with the SM. The fit result is:
\begin{eqnarray}
\tfrac{\BRof{\bs}}{\BRSMof{\bs}} = 0.76^{+0.20}_{-0.18}  \,,\nonumber \\
\tfrac{\BRof{\bd}}{\BRSMof{\bd}} = 3.7^{+1.6}_{-1.4}  \,. \nonumber 
\end{eqnarray}
The compatibility of \BRof{\bsmumu} and \BRof{\bdmumu} with the SM is evaluated to be $1.2\,\sigma$ and $2.2\,\sigma$ respectively. These numbers also take the theoretical uncertainties into account. 

\begin{figure}[!h]
\centering
\includegraphics[height=2in]{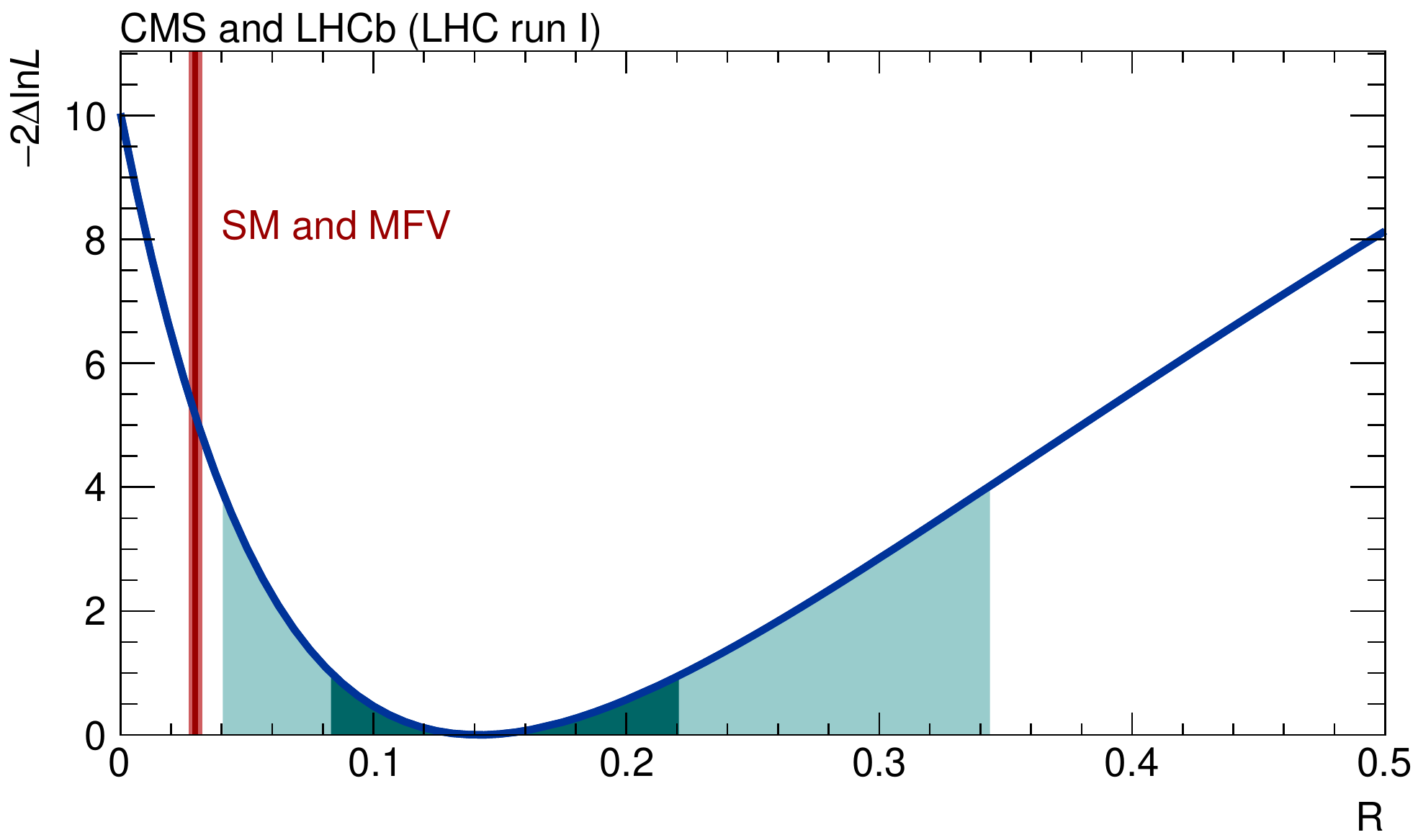}
\caption{Likelihood test-statistics ($-2\Delta \mathrm{ln}L$) profile for \RB. The dark and light (cadet blue) areas define the $\pm 1\,\sigma$ and $\pm 2\,\sigma$ confidence intervals, respectively. The SM expectation and its uncertainty for each BF is denoted with a vertical (red) band.}
\label{fig3}
\end{figure}

A separate fit to the ratio of \bd to \bs gives $\RB = 0.14^{+0.08}_{-0.06}$, which is compatible with the SM of $2.3\,\sigma$, including the theoretical errors. The likelihood profile for \RB is shown in Fig.~\ref{fig3}.
%%%%%%%%%%%%%%%%%%%%%%%%%%%%%%%%%%%%%%%%%%%%%%%%%%%%%%%%%%%%%%%%%%%%%%%%%
%%%%%%%%%%%%%%%%%%%%%%%%%%%%%%%%%%%%%%%%%%%%%%%%%%%%%%%%%%%%%%%%%%%%%%%%%%%

\section{Conclusions}

The combination of \lhcb and \cms results on \bdsmumu searches exploiting the full statistics of Run I at LHC has been presented. 
The \bsmumu decay is observed for the first time with $6.2\,\sigma$ statistical significance, with a BF compatible with the SM within $1.2\,\sigma$. A  $3.0\,\sigma$ excess is observed for the \bdmumu with respect to the background-only hypothesis. The compatibility of this channel with the SM is measured to be $2.2\,\sigma$. 
The ratio of \bd to \bs BFs, \RB is compatible with the SM within $2.3\,\sigma$. The ATLAS measurement of \bsmumu is not mentioned in these proceedings, since an update of the analysis of 2011 data using the full dataset will hopefully be available soon.

\end{document}